# Optically induced lattice deformations, electronic structure changes, and enhanced superconductivity in $YBa_2Cu_3O_{6.48}$

R. Mankowsky[1], M. Fechner[1], M. Först[1], A. von Hoegen[1], J. Porras[2], T. Loew[2], G.L. Dakovski[3], M. Seaberg[3], S. Möller[3], G. Coslovich[3], B. Keimer[2], S.S. Dhesi[4], and A. Cavalleri[1,5]

[1]*Max Planck Institute for the Structure and Dynamics of Matter, Hamburg, Germany*

[2]*Max Planck Institute for Solid State Research, Stuttgart, Germany*

[3]*Linac Coherent Light Source, Stanford Linear Accelerator Center (SLAC) National Accelerator Laboratory, Menlo Park, CA*

[4] *Diamond Light Source, Harwell Science and Innovation Campus, Didcot OX11 0DE, United Kingdom*

[5]*Department of Physics, Clarendon Laboratory, University of Oxford, UK*

## ABSTRACT

Resonant optical excitation of apical oxygen vibrational modes in the normal state of underdoped $YBa_2Cu_3O_{6+x}$ induces a transient state with optical properties similar to those of the equilibrium superconducting state. Amongst these, a divergent imaginary conductivity and a plasma edge are transiently observed in the photo-stimulated state. Femtosecond hard x-ray diffraction experiments have been used in the past to identify the transient crystal structure in this non-equilibrium state. Here, we start from these crystallographic features and theoretically predict the corresponding electronic rearrangements that accompany these structural deformations. Using density functional theory, we predict enhanced hole-doping of the $CuO_2$ planes. The empty chain Cu $dy^2$-$z^2$ orbital is calculated to strongly reduce in energy, which would increase c-axis transport and potentially enhance the interlayer Josephson coupling as observed in the THz-frequency response.
From these results, we calculate changes in the soft x-ray absorption spectra at the Cu $L$-edge. Femtosecond x-ray pulses from a free electron laser are used to probe changes in absorption at two photon energies along this spectrum, and provide data consistent with these predictions.

**Introduction**

Unconventional high-temperature superconductivity in the cuprates appears as holes are doped into the $CuO_2$ planes of the parent compounds. $YBa_2Cu_3O_{6+x}$ is a bilayer superconductor, made up of two $CuO_2$ planes per unit cell, stacked along the crystal *c*-axis. The hole doping of these planes is controlled by the oxygen content of Cu-O chains, which form along the along the *b*-axis in between the bilayers. A superconducting phase appears with a maximum transition temperature $T_C$ of 93 K at optimal doping with x = 0.92. The increase in $T_C$ with hole doping is accompanied by a decrease in the distance between the apical oxygen O(4) and the planar copper atoms Cu(2) [1].

In the underdoped regime, the critical temperature can further be enhanced by the application of static pressure [2,3,4,5]. This effect has been attributed to an increase in hole doping as a result of additional electron transfer from the $CuO_2$ planes to the Cu-O chains, driven by a reduction in the Cu(2)–O(4) distance [6]. However, the application of pressure does not simply correspond to a movement along the doping coordinate. In fact, the critical temperature of underdoped compounds under pressure can exceed that at optimal doping.

**Optically induced lattice deformations and enhanced superconductivity**

Resonant optical excitation of vibrational modes by intense mid-infrared light pulses has recently emerged as a new way to control the superconducting state away from equilibrium, creating transient crystal structures [7,8] and enhancing superconducting coherence in many different compounds [9,10,11]. In $YBa_2Cu_3O_{6.5}$, the large-amplitude excitation of the *c*-axis apical oxygen-copper stretching mode at oxygen vacant chain sites has been shown to induce coherent interlayer transport up to room temperature [12,13]. The key results of these time-resolved THz spectroscopy experiment are summarized in Figure 1. The equilibrium superconducting state of $YBa_2Cu_3O_{6.5}$ at 10 K is characterized by a $1/\omega$ divergence for $\omega \to 0$ in the imaginary part of the optical conductivity $\sigma_2$ and the Josephson plasma edge in the THz reflectivity, a direct result of coherent interlayer tunnelling (Fig. 1a,b). The light-induced changes of these properties 0.8 ps after vibrational excitation of $YBa_2Cu_3O_{6.5}$ at 100 K, well above the transition temperature of 55 K, are shown in the lower panels (data taken from [12]).

The transient properties exhibit striking similarities to the equilibrium superconducting state, featuring both the plasma edge in the reflectivity and the $1/\omega$ divergence of $\sigma_2$ toward lower frequencies (dashed lines). This effect is most pronounced in underdoped compounds and vanishes at optimally doped $YBa_2Cu_3O_{6.92}$, with signatures of light-induced coherent tunneling found up to 300 K in $YBa_2Cu_3O_{6.5}$.

The transient lattice structure of $YBa_2Cu_3O_{6.5}$, as determined by femtosecond x-ray diffraction experiments at 100 K [14], is shown in Figure 2a. Three key effects were identified. First, the buckling of the O-Cu-O bonds in the $CuO_2$ planes was increased both at oxygen vacant and filled chain sites. Secondly, the $CuO_2$ bi-layers were driven away from one another, while the inter-bilayer distance reduced. Thirdly, the Cu(2) – O(4) distance *d* at oxygen vacant chain sites was reduced by around 1%. The temporal evolution of these atomic rearrangements was extracted from the recorded diffraction intensity, shown for two different Bragg reflections in Figure 2b.

Figure 2c shows the time-resolved change in apical oxygen-planar Cu distance $d$. After excitation, this distance reduced, recovering with the same temporal profile as the optical signatures of superconductivity.

**Changes of the electronic structure**

Starting from these results, we calculated the partial density of states (DOS) of all Cu atoms for the equilibrium and the transient structure, as acquired from a projection of the total density of states onto their muffin tin spheres (see computational details section at the end of the manuscript). Note that in ortho-II ordered $YBa_2Cu_3O_{6.48}$, the Cu-O chains are alternatingly filled and vacant of oxygen atoms. Thus, the unit cell contains four distinct copper sites, the $Cu(1)_{v,f}$ atoms at oxygen vacant ($v$) and filled ($f$) chain sites, as well as the adjacent $Cu(2)_{v,f}$ atoms in the $CuO_2$ planes, as illustrated in Figures 2 and 3.

Figure 4 presents the partial DOS of the $d_{y^2-z^2}$ orbital of the chain and the $d_{x^2-y^2}$ and $d_{z^2}$ orbitals of the planar copper atoms, both at oxygen vacant and filled chain sites for the equilibrium (grey) and displaced structure (red). The results are shown here exemplarily for an excitation amplitude corresponding to a reduction in Cu(2) – O(4) distance $d$ by 10%. The largest changes in the partial DOS are associated with the decrease in apical-O planar Cu distance and are found at the $Cu(1)_v$ oxygen vacant chain site, where the unoccupied $d_{y^2-z^2}$ density significantly downshifts in energy. This change is accompanied by an energy increase of the occupied $dz^2$ density at the adjacent planar Cu(2)$_v$ site (Fig. 4a–d).

We also analyzed the $dx^2$-$y^2$ orbitals of the Cu(2) sites, aligned in the CuO$_2$ planes, which are mainly affected by the increase in O-Cu-O bond angle due to the reduced hybridization with the neighbouring oxygen $p$-orbitals. However, as the density of the Cu(2)$_{v,f}$ $dx^2$-$y^2$ orbitals exhibited only minor changes, both next to oxygen vacant and oxygen filled chain sites, we conclude that the increase in buckling is unlikely to induce any charge transfer dynamics (Fig. 4e,f).

Figure 5a shows the relative changes in the occupation of the Cu $d$-orbitals, calculated from the structurally driven changes in the densities of states at different amplitudes. We found an increase in occupation of the chain Cu(1)$_v$ $dy^2$-$z^2$ and planar Cu(2)$_v$ $dx^2$-$y^2$ orbitals at oxygen vacant sites. At filled sites, the chain Cu(1)$_f$ $dy^2$-$z^2$ is unaffected while the occupancy of the Cu(2)$_v$ $dx^2$-$y^2$ reduces. The structural rearrangement thus lifts the degeneracy in occupation between the $dx^2$-$y^2$ orbitals at empty and filled chain sites. Note that on average, the occupation of the planar $dx^2$-$y^2$ orbitals reduces, indicating an increase in hole doping.

Experimentally, the changes in the density of states can be measured by x-ray absorption spectroscopy at the Cu $L$-edge, which probes transitions from the Cu $2p$ core levels to the unoccupied $d$-orbitals. Figure 5b shows the calculated absorption spectrum for x-rays polarized along the crystal $c$-axis ($E//c$), sensitive to transitions into the Cu(1) $dy^2$-$z^2$ orbitals. The spectrum features three main peaks. The two peaks at lower energy originate from transitions into the unoccupied DOS of the filled chain $Cu(1)_f$ atom. The third peak results from the monovalent oxygen vacant chain $Cu(1)_v$ atom [15,16,17].

As a consequence, the energy reduction for the $dy^2$-$z^2$ empty-chain $Cu(1)_v$-density appears in the $E//c$ XAS spectrum as spectral weight transfer from the highest energy peak toward lower energies. The corresponding relative changes of the XAS spectrum are shown in Figure 5c, calculated for different amounts of lattice distortion.

Figure 6a shows the experimental $E//c$ XAS spectra, measured at the Cu-$L$ edge of detwinned $YBa_2Cu_3O_{6.48}$ samples (with hole doping of p = 0.095) at 100K [18], well above the transition temperature of $T_C$ = 55 K. The data were acquired by detecting the sample total fluorescence yield from synchrotron radiation at the Diamond Light Source, UK. The measured and calculated spectra show excellent agreement, however with the calculated spectrum being slightly broader in energy.

To verify the calculated charge redistribution in the transient crystal structure and to follow its temporal evolution, we measured time-resolved changes in the XAS spectrum induced by the same 4mJ/cm² 15-μm mid-IR excitation of the experiments reported above. These experiments were carried out using femtosecond x-ray pulses from the LCLS free electron laser [19] on the same sample and at 100 K temperature.

Figure 6b shows the measured relative changes in $YBa_2Cu_3O_{6.48}$ fluorescence intensity at photon energies of 932 eV (blue) and 933.2 eV (red), corresponding to the second and third peak of the experimental XAS spectrum. The solid lines are fits to the data with recovery times fixed to those of the transient lattice rearrangement and the optical signatures of light-induced superconductivity, measured by time-resolved x-ray diffraction [14] and THz spectroscopy [12,13], respectively. Hence, only the amplitude is left as free fitting parameter.

The intensities at the two peaks in the XAS was found to change in opposite directions, increasing by +0.5% at the lower energy $Cu(1)_f$ peak and reducing by –0.4% at the at the higher energy oxygen vacant $Cu(1)_v$ peak. Figure 6c shows the calculated changes in XAS intensity at the two peaks for different amplitudes of lattice distortion (solid lines) together with the experimental data (markers). These are consistent in both sign and magnitude for a structural rearrangement corresponding to a change in Cu(2)-O(4) distance of 0.2%, which, however, is smaller than $\Delta d/d$ of 1% extracted from the x-ray diffraction measurements (Ref. 14). This deviation might arise either from an overestimation of the excitation fluence in the experiment or from uncertainties in the calculations, as for example the value of the onsite Coulomb repulsion U (see computational details).

We now discuss some possible implications of these charge transfer dynamics on superconductivity. First, in analogy with the pressure-induced charge transfer dynamics the effective self-doping of the planes could enhance the transition temperature. The total increase in doping is approximately 0.01 holes per $(CuO_2)_v$ unit for a change in apical oxygen-copper distance of 1% measured by femtosecond x-ray diffraction. This value is comparable to pressure induced effects ranging from 0.005 to 0.02 holes/GPa per $(CuO_2)$ unit [6, 20, 21, 22]. From the pressure-dependence of the transition temperature [2-6], we estimate an enhancement of $T_C$ on the order of (3-12) K. Despite the similarities to pressure induced effects, the lifting of the degeneracy in occupancy of the two planar Cu(2) atoms at oxygen vacant and filled chain sites is unique to mid-infrared excitation.

Secondly, we identified a large increase in the orbital occupancy of the empty chain Cu(1)$_v$ $dy^2$-$z^2$ orbitals. Together with the increase in orbital overlap with the planar Cu(2)$_v$ $dz^2$ orbital, the *c*-axis transport properties could be strongly modified. While at this point we can only speculate on the connection of these two effects to the superconducting properties, we note that the time-resolved THz spectroscopy experiments have indeed found an increase in *c*-axis Josephson coupling of adjacent CuO$_2$ planes through the interlayer region.

**Summary**

To summarize, we analyzed the electronic dynamics following the same resonant lattice excitation of YBa$_2$Cu$_3$O$_{6.48}$ that has previously been shown to induce transient coherent transport above $T_C$ and to deform the materials crystal structure. We used density functional theory to calculate the electronic rearrangement driven by this lattice deformation. These calculations predicted an enhancement in hole doping of the CuO$_2$ planes, accompanied by an increase in orbital overlap along the crystal *c*-axis. We validated these predictions by measuring changes in x-ray absorption at the Cu *L*-edge with femtosecond resolution. The experimental data revealed that the charge transfer dynamics have the same temporal profile as the signals of light-induced coherent transport and the structural deformation, indicating an intimate connection.

**FIGURES**

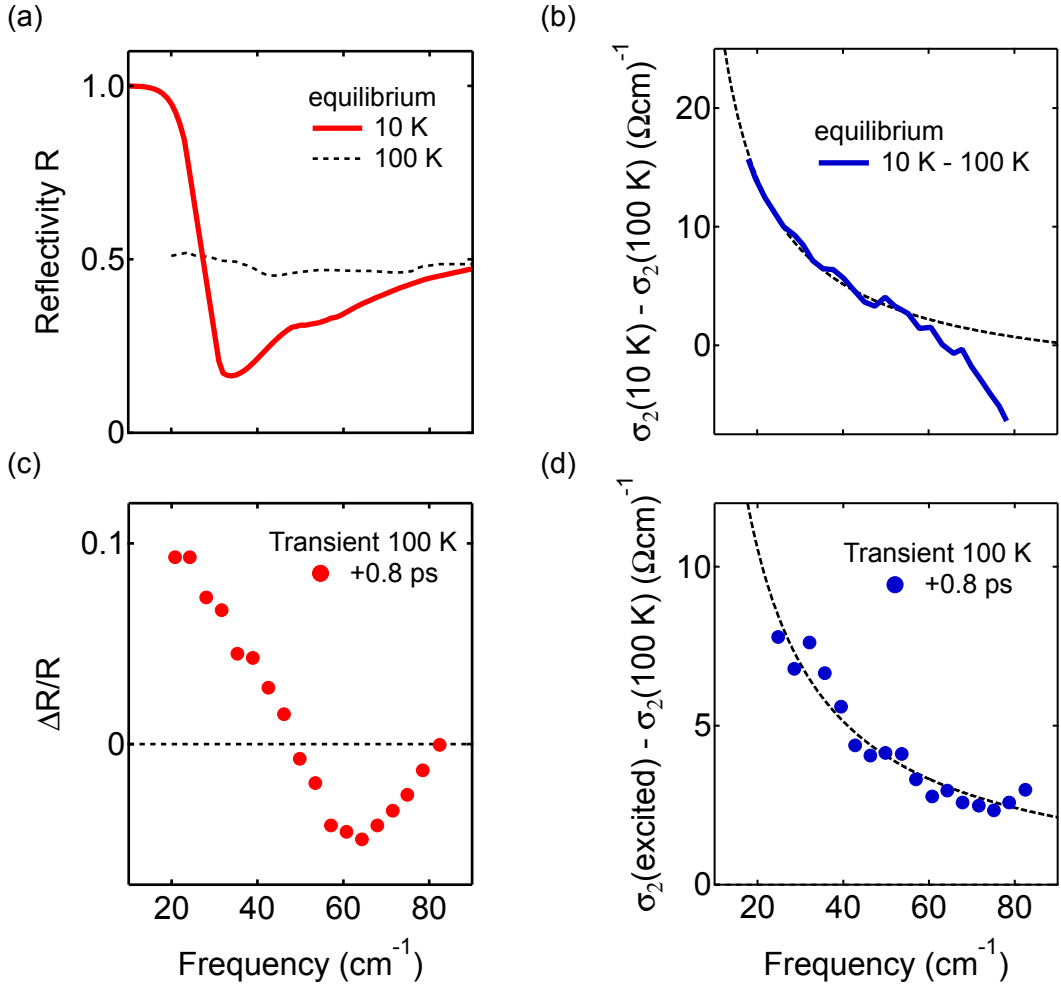

**Figure 1.** (a) Equilibrium THz reflectivity of $YBa_2Cu_3O_{6.5}$ above (dashed line) and below the transition temperature of $T_C$ = 55 K (solid line), showing the appearance of an edge. (b) The temperature-driven change in the imaginary part of the optical conductivity $\sigma_2$ (solid line) across the metal-superconductor transition exhibits a $1/\omega$ divergence for $\omega \to 0$ (dashed line). (c,d) Light-induced changes in these properties, 0.8 ps after vibrational excitation of $YBa_2Cu_3O_{6.5}$ at 100 K with 300-fs, 15-μm mid-infrared pulses, show signatures similar to the equilibrium superconducting state (from [12]).

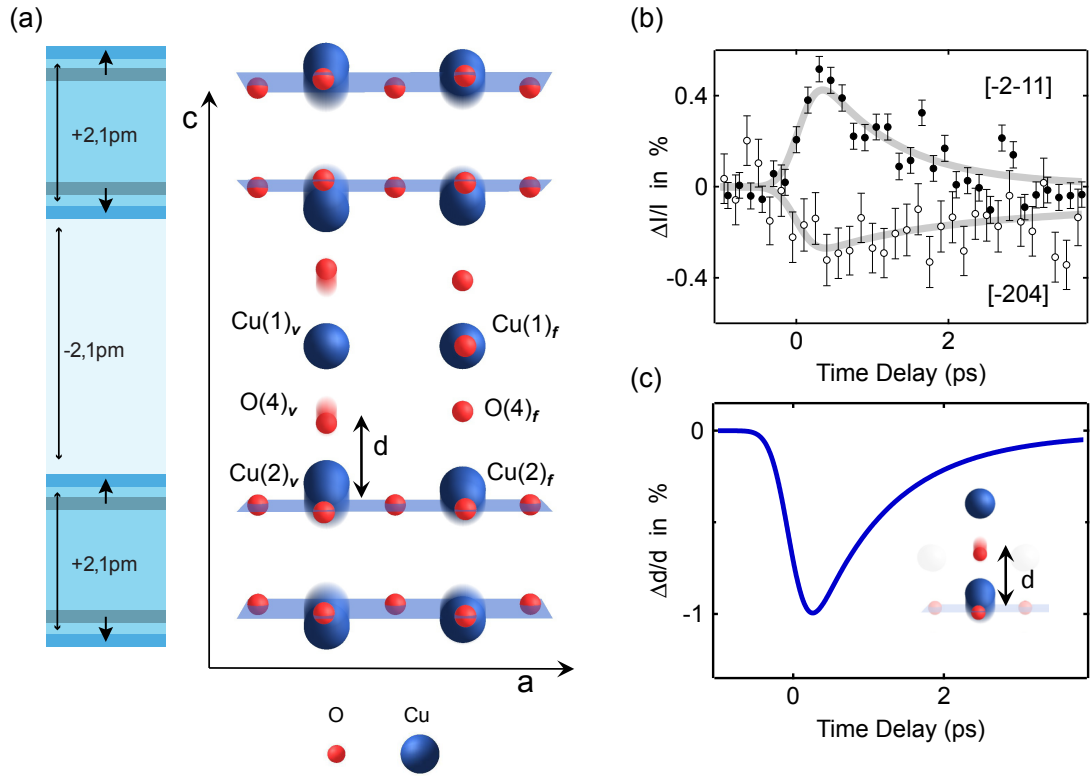

**Figure 2.** (a) Crystal structure of orthorhombic $YBa_2Cu_3O_{6.5}$, showing the apical oxygen O(4) atoms and the copper atoms in the $CuO_2$ planes $Cu(2)_{v,f}$ and the Cu-O chains $Cu(1)_{v,f}$, which form along the *b*-axis, both at oxygen vacant and filled chain sites (*v, f*). For simplicity, only Cu and O atoms are shown. In the light-induced superconducting state, driven by resonantly exciting the *c*-axis $B_{1u}$ phonon mode, the distance between the apical oxygen atoms (red shadows) and the planar copper atoms (blue shadows) decreases. (b) Changes in x-ray diffraction intensity recorded at two different Bragg reflections following excitation of $YBa_2Cu_3O_{6.5}$ at 100 K. Error bars are 1σ (67% confidence interval). (c) Temporal evolution of the change in $O(4)_v$ - $Cu(2)_v$ distance *d* (from [14]).

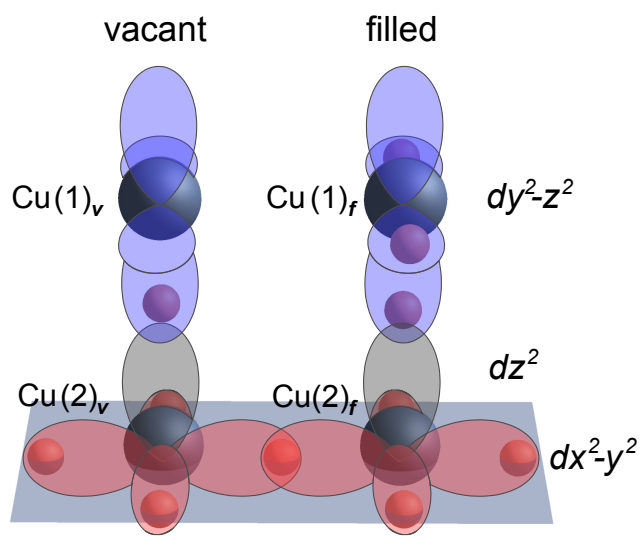

**Figure 3:** Sketch of the $dy^2$-$z^2$ orbitals of the chain Cu(1) atoms (blue) and $dx^2$-$y^2$ orbitals of the planar Cu(2) atoms (red) at filled and oxygen vacant chain sites. The $dz^2$ orbitals of the planar Cu(2) atoms are shown in grey.

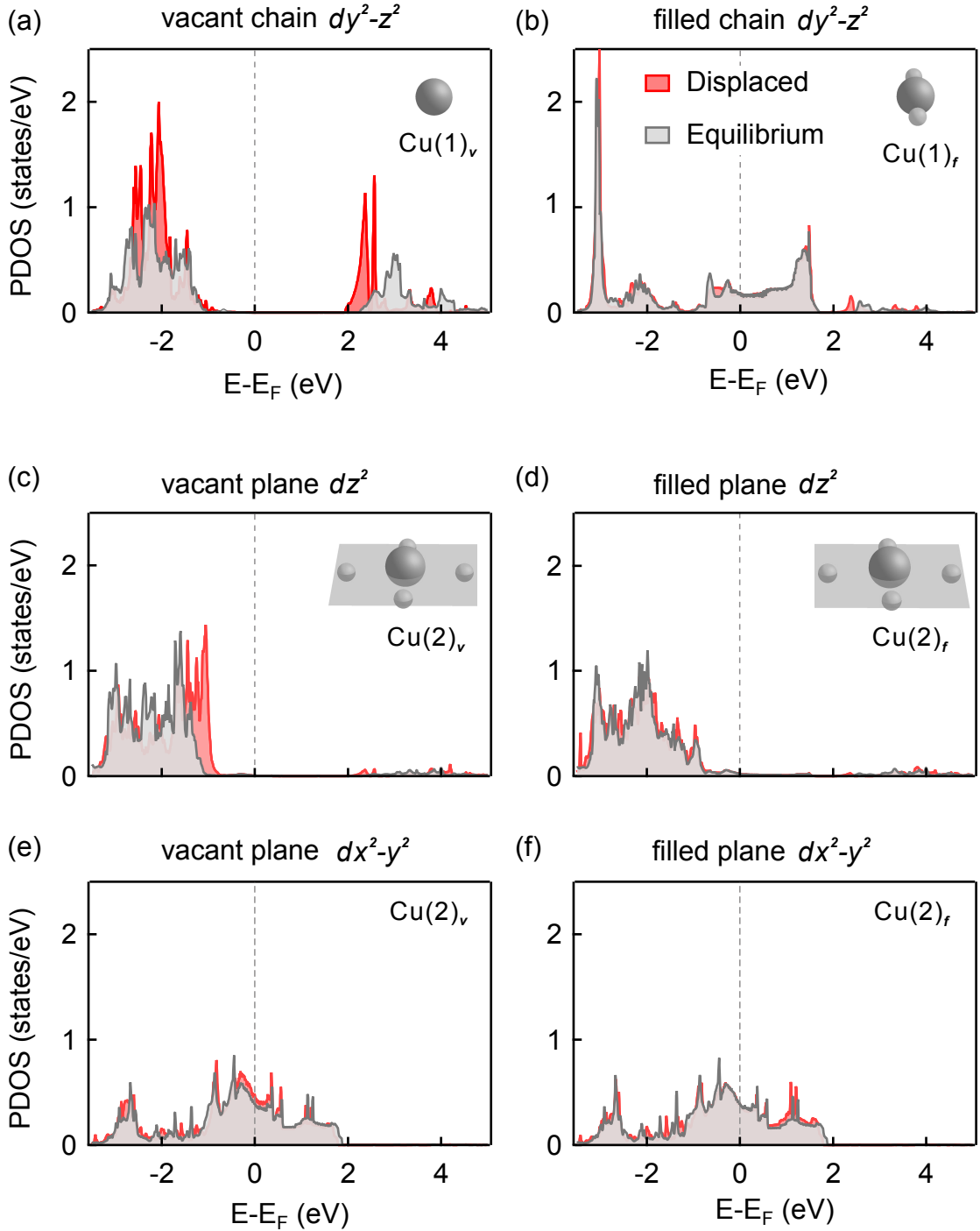

**Figure 4:** (a-f) Projected partial density of *d*-states for different copper sites within the YBa$_2$Cu$_3$O$_{6.5}$ structure. Grey and red lines show the density of states of the ground state and the transient crystal structure for a displacement corresponding to a change in Cu(2)-O(4) distance of Δd/d = 10%, respectively.

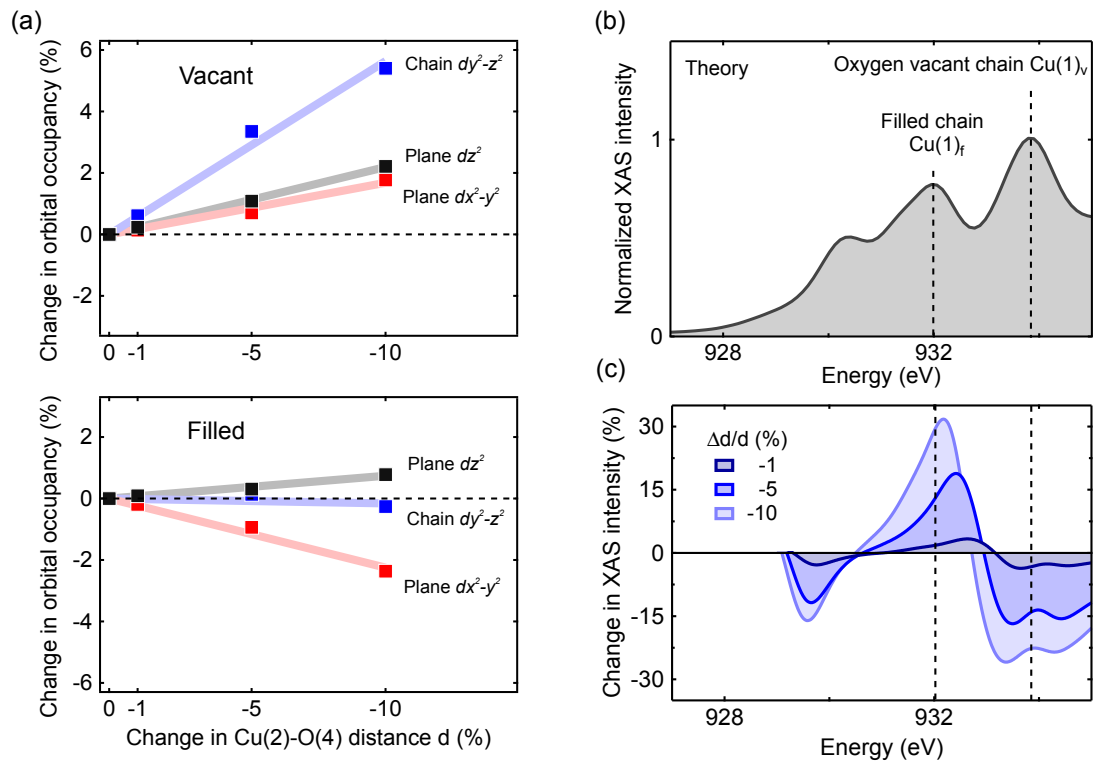

**Figure 5:** (a) Change in orbital occupation of the chain Cu(1) $dy^2$-$z^2$ and planar Cu(2) $dx^2$-$y^2$ and $dz^2$ orbitals for different amplitudes of the structural modulation driven by phonon excitation both at oxygen vacant (upper panel) and filled chain sites (lower panel). (b) Computed x-ray absorption spectrum (XAS) of $YBa_2Cu_3O_{6.5}$. The vertical lines mark the peaks related to the oxygen vacant and filled Cu(1) chain sites. (c) Modulations in the XAS spectrum due to different amounts of structural distortions shown as relative differences with respect the equilibrium spectrum.

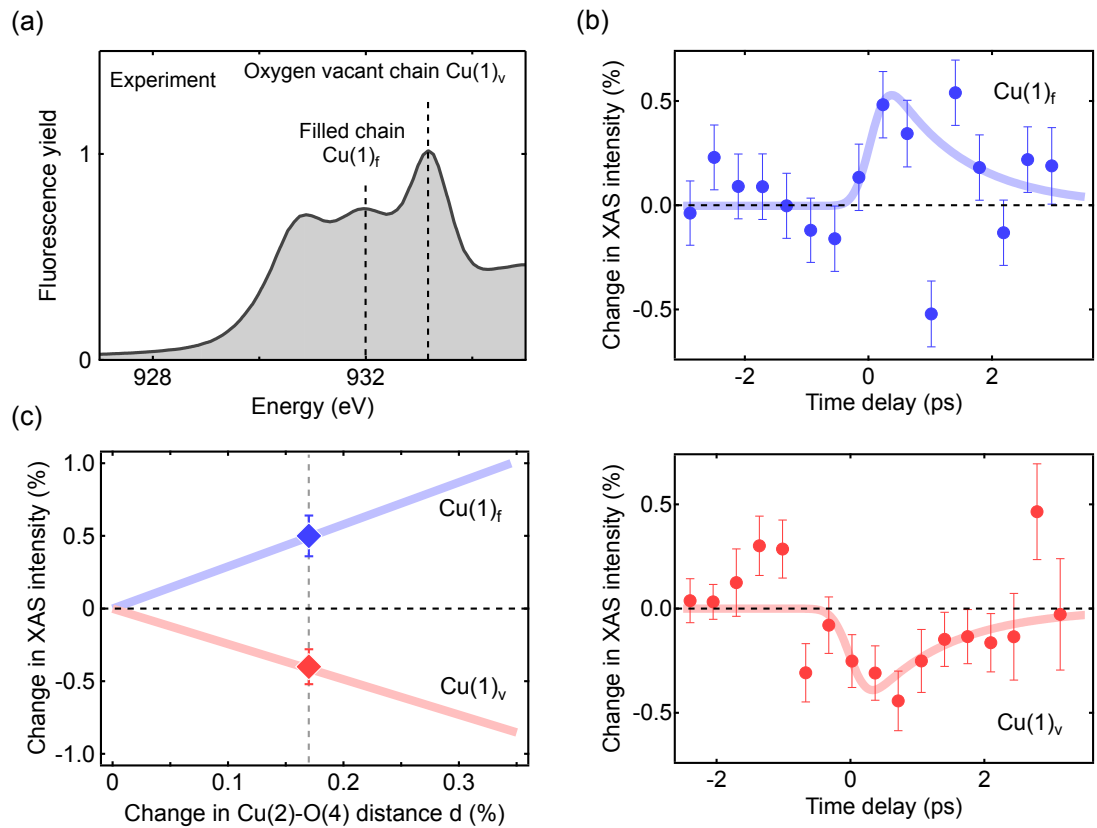

**Figure 6.** (a) Equilibrium XAS spectrum of YBa$_2$Cu$_3$O$_{6.48}$ at 100 K, acquired by measuring the fluorescence yield. The dashed lines indicate the two peaks sensitive to the occupation of the chain copper atoms at filled Cu(1)$_f$ and vacant Cu(1)$_v$ chain sites. (b) Time-resolved changes in fluorescence intensity of these peaks, Cu(1)$_f$ at 932 eV in blue and Cu(1)$_v$ at 933.2 eV in red, induced by the resonant lattice excitation of the *c*-axis B$_{1u}$ phonon mode with 150-fs mid-IR pulses at 15 μm wavelength and a fluence of 4 mJ/cm$^2$. The solid lines are fits to the data (see text). Error bars are 1σ (67% confidence interval). (c) Calculated changes of the XAS intensity for different amplitude in lattice rearrangement (solid lines) and experimental values obtained from the fit to the data of panel b (Markers).

**Computational Details**

To compute the electronic structure and x-ray absorption spectra of $YBa_2Cu_3O_{6.5}$ we performed first principles calculations within the framework of density functional theory. Our computational approach based on the full potential linearized augmented plane waves (FLAPW) method as implemented in the ELK code [23]. All calculations were performed for experimental structures given in Ref. [14]. We used the generalized gradient approximation by Perdew-Burke-Ernzerhof (PBE)[24] for the exchange-correlation functional and included spin-orbit coupling by the second variational method. Owing to correlated Cu-3d electrons, we additionally use the plus Hubbard *U* approach of Lichtenstein et al. [25] with a U=3 eV and J=0 eV. The specific *U* choice has been made since it reproduces the best agreement of the computed XAS spectrum with the static measured spectra. Please note, that we apply no +U at $Cu(1)_v$ sites since its nominally $Cu^{1+}$ configuration exhibits a full d-shell.

The numerical parameters of our computations are a truncation at $l_{max}$=10 of the angular expansion of wave functions and potential within the muffin-tin radii (2.6, 2.8, 1.85 and 1.4 a.u. for Y, Ba, Cu and O, respectively). The potential and density expansion within the interstitial region is limited by $|G|max=20$ a.u.$^{-1}$ and the plane-wave cutoff of the wavefunctions is set to $R_{MT}$ x $k_{max}$=8.0. The Brillouin zone is sampled within our computations by a 11x19x5 *k-point* mesh. We compute the X-ray absorption spectra within the Kubo linear response formalism as described in Ref. [26], where in our calculation we explicitly treat Cu 2-*p* states as valence electrons. All theoretical spectra were broadened with a 0.3 eV wide (FWHM) Lorentzian to account for lifetime broadening.

To trace the amount of charges/holes at each atomic site, we follow the method of Ref. [27,28] and project the charges within each muffin-tin sphere onto site symmetrized local orbitals. We note, that while the absolute amount of holes cannot be assigned by this approach, the calculations still yield a good estimate of relative changes even if the muffin-tin spheres exhibit a limited coverage of total volume.

**Experimental Details**

The equilibrium XAS measurements were performed on beamline I06 at the Diamond Light Source, UK. The time-resolved XAS experiments were carried out at the LCLS x-ray free electron laser. Mid-infrared pulses of 300 fs duration and 4 mJ/cm² fluence at 15 μm wavelength were used to resonantly drive the 670-cm$^{-1}$ infrared-active apical oxygen phonon mode polarized along the crystal *c* axis. The same excitation was demonstrated to induce coherent *c*-axis interlayer coupling in underdoped YBa$_2$Cu$_3$O$_{6+\delta}$ compounds above $T_C$, reminiscent of superconductivity [12,13]. The YBa$_2$Cu$_3$O$_{6.48}$ sample was mounted in a vacuum chamber and cooled to 100 K, well above its critical temperature of 55 K. X ray pulses tuned to the Cu L-edge and monochromatized to 0.5-eV bandwidth were used to measure the time-dependent transient fluorescence intensity, which was detected by a multi-channel plate (MCP) detector. Pulse-to-pulse intensity normalization to the incident x-rays was facilitated by a second MCP detector upstream of the experiment.

**ACKNOWLEDGEMENT**

Portions of this research were carried out on the SXR Instrument at the Linac Coherent Light Source (LCLS), a division of SLAC National Accelerator Laboratory and an Office of Science user facility operated by Stanford University for the U.S. Department of Energy. The SXR Instrument is funded by a consortium whose membership includes the LCLS, Stanford University through the Stanford Institute for Materials Energy Sciences (SIMES), Lawrence Berkeley National Laboratory (LBNL, contract No. DE-AC02-05CH11231), University of Hamburg through the BMBF priority program FSP 301, and the Center for Free Electron Laser Science (CFEL). Also, we acknowledge the provision of beamtime by Diamond Light Source at the I06 beamline under proposal number SI13341.

The research leading to these results has received funding from the European Research Council under the European Union's Seventh Framework Programme (FP7/2007-2013) / ERC Grant Agreement n° 319286 (Q-MAC).

This work has been supported by the excellence cluster 'The Hamburg Centre for Ultrafast Imaging - Structure, Dynamics and Control of Matter at the Atomic Scale' of the Deutsche Forschungsgemeinschaft.